# Conditions for the growth of smooth La$_{0.7}$Sr$_{0.3}$MnO$_3$ thin films by pulsed electron ablation

P. Graziosi [a,b,*], M. Prezioso [b], A. Gambardella [b], C. Kitts [b,1], R.K. Rakshit [b,2], A. Riminucci [b], I. Bergenti [b], F. Borgatti [b], C. Pernechele [c], M. Solzi [c]. D. Pullini [d], D. Busquets-Mataix [a,e], V. A. Dediu [b]

[a] Instituto de Tecnología de Materiales, Universidad Politécnica de Valencia, Camino de Vera s/n, 46022, Valencia, Spain

[b] CNR - ISMN, Consiglio Nazionale delle Ricerche - Istituto per lo Studio dei Materiali Nanostrutturati, v. Gobetti 101, 40129, Bologna, Italy

[c] Dipartimento di Fisica, Università di Parma, Area delle Scienze 7/A, 43124, Parma, Italy

[d] Centro Ricerche Fiat, 10043, Orbassano (TO), Italy

[e] Departamento de Ingeniería Mecánica y de Materiales, Universidad Politécnica de Valencia, Camino de Vera s/n, 46022, Valencia, Spain

[1] Present address: Department of Physical Chemistry, Chalmers University of Technology, 41296, Gothenburg, Sweden

[2] Present address: Quantum Phenomena and Applications Division, National Physical Laboratory, Council of Scientific and Industrial Research, Dr. K. S. Krishnan Marg, New Delhi 110012, India

[*] Corresponding author: patrizio.graziosi@gmail.com, tel. +39 0516398505, +39 0516398517, fax +39 0516398540

**Abstract**

We report on the optimisation of the growth conditions of manganite $La_{0.7}Sr_{0.3}MnO_3$ (LSMO) thin films prepared by Channel Spark Ablation (CSA). CSA belongs to pulsed electron deposition methods and its energetic and deposition parameters are quite similar to those of pulsed laser deposition. The method has been already proven to provide manganite films with good magnetic properties, but the films were generally relatively rough (a few nm coarseness). Here we show that increasing the oxygen deposition pressure with respect to previously used regimes, reduces the surface roughness down to unit cell size while maintaining a robust magnetism. We analyse in detail the effect of other deposition parameters, like accelerating voltage, discharging energy, and temperature and provide on this basis a set of optimal conditions for the growth of atomically flat films. The thicknesses for which atomically flat surface was achieved is as high as about 10-20 nm, corresponding to films with room temperature magnetism. We believe such magnetic layers represent appealing and suitable electrodes for various spintronic devices.

1. Introduction

Manganese perovskites are among the most employed laboratory materials for various spintronic devices [1]. Impressive efforts have been made in order to understand the complex physics characteristic for these materials, implicating massive experimental and theoretical investigations [2, 3]. In addition the interest towards device applications stimulated detailed surface studies including a wide spectrum of various spectroscopy [4] and microscopy techniques [5], supported often by thorough *ab initio* simulations [6, 7].

A substantial revival of the interest towards manganites has been stimulated recently by the demonstration of the extremely efficient utilization of $La_{0.7}Sr_{0.3}MnO_3$ as spin polarized electrode in various spintronic devices with organic semiconductors spacers in both tunnelling [8] and injection regimes [9, 10]. Although the commercial application of this and other manganites is still strongly questioned by their relatively low Curie temperature (about 350-370K) [11], they represent an excellent tool for unravelling most basic and important effects in spintronic systems involving organic semiconductors [8, 12, 13]. Such application advances nevertheless extremely tough requirements to the sharpness of all the involved interfaces, and hence to the smoothness of all the components surfaces. For example, an interface roughness exceeding 1-2 nm (approximately molecular size) seriously questions the reproducibility of main spintronic effects [14]. This pushes the investigators to develop atomically flat device components combined in as sharp as possible multilayered assembly.

Ablation techniques have proven so far to be extremely efficient in the deposition of thin films of complex oxides, like High-$T_C$ superconductors (HTC), ferroelectric oxides and colossal magnetoresistive manganites (CMR) [15]. Among different ablation techniques, several modifications of the pulsed electron deposition (PED) method are recently successfully competing in the deposition of oxide superconductors and manganites with the most widespread ablation technique - Pulsed Laser Deposition (PLD). Thus the Channel Spark Ablation (CSA) [16] has proved the deposition of high quality thin films of simple oxides, such as $Al_2O_3$ [17, 18], and complex oxides such as cuprates [19] and manganites [20]. Previous X-ray Diffraction and Raman investigations on selected samples of different thicknesses have ALWAYS indicated (001) growth orientation on both STO (100) and NGO (110). This paper is dedicated to morphology and we did not perform sample by sample XRD characterizations, such characterizations together with Transmission Electron Microscopy (TEM) analysis are currently in progress and will be published elsewhere.

Especially successful was the application of CSA to LSMO films, where excellent magnetic, electric and structural properties have been achieved [20-22]. Moreover, manganite electrodes grown by CSA have so far shown best performances in organic spintronics including highest achieved tunnelling magnetoresistance (TMR) exceeding 300% [8] and room temperature operation of injection devices [9, 23].

The surface roughness of these films was nevertheless relatively high and also the thinnest LSMO films (below 10 nm) presented a rms roughness of a few nanometers due to the presence of particulate and/or outgrowths [21].

In order to compare state of the art roughnesses for films produced by CSA and other deposition methods we describe below selected cases for few mainly involved techniques. The lowest roughness achieved by an ablation technique was performed by PLD and found to be about 0.2 nm on about 10 nm thick films [24]. Similar results were featured by Metalorganic Aerosol Deposition (MAD) [25, 26], but in that case the 0.2 nm roughness is claimed for thicker films of 80 nm. Best MBE roughness reached as low as 0.4 nm [27] and was found for 9 nm films. Noteworthy the lowest roughness presented in literature was achieved by sputtering technique and was represented by 0.11 nm value for 7 nm thick films [28]. Unfortunately in the latter case the films of this distinguished flatness were not ferromagnetic at room temperature – moreover, RT magnetism was achieved only for films exceeding 27 nm thickness.

With this picture in mind and considering exceptional results achieved by CSA manganites in organic spintronic devices, the ability of CSA to reach roughness values comparable to best reported ones represents the main goal and motivation of this paper. We will concentrate our efforts on very thin films, which, once achieved stable room temperature magnetism, offer most versatile electrode performances, like for example homogeneous (throughout the thickness) strain engineering, oxygen control, low density of defects etc. . Very thin films (10-20 nm and below) are much less investigated but they are considered as most appropriate for spintronic applications and a deep investigation in order to achieve very thin film with robust magnetotransport properties is going on [24]. Working with very thin films is thus one of the explicit goals of our paper. We will show indeed that 0.2 nm roughness can be achieved by CSA technique for thicknesses suitable for the described above applications.

2. Experimental details

The CSA system is depicted in Figure 1. The principle consists in the ablation of a stoichiometric target by means of a pulsed electron beam [29]. A plasma is triggered in order to extract electrons for the beam. A negative high voltage (5-30 KV) power supply is directly connected to a hollow cathode (a) and a capacitor (b). The latter is grounded through an air gap (c) having a floating electrode which is decoupled from the capacitor by a charging resistor (d). Between the charging resistor (d) and the floating electrode of the air gap a triggering anode plate (e) is located and inserted in the bulb (f). At a sufficient high voltage a spark brakes down the air gap (c), a rapid variation of the electric field between the hollow cathode and the anode plate ionizes the gas molecules in the bulb triggering plasma in the cathode cavity (a), where the amplification of the discharge happens. Because of the high resistance of the charging resistor, the capacitor discharge happens through the low impedance electron beam (g). The negative high voltage accelerates the electrons of the beam in the Pyrex channel (h), where the beam is further amplified by secondary emission from the channel wall. The active role of the channel gives the technique its name. The electron beam current would be continuous if the power supply could provide a sufficient current, but since it is limited in current the electron beam cannot be sustained and the discharge extinguishes up to the new spark in the air gap. As a consequence the beam assumes a pulsed character. The current supplied to charge the capacitors defines the charging time and, hence, the operating frequency. The high voltage and the capacitance determine the accumulated charge and the total energy. The energy distribution of the electrons in the beam and the length of the pulse is determined by the accelerating voltage and the gas pressure. Since the environment gas "stuffs" the beam, which is space charge neutralized the gas acts like a "window" which the electrons pass through and by the pressure one controls the electron energy distribution in the beam. As a consequence only a narrow range of pressures is able to stabilize the beam, from 1 to $5 \cdot 10^{-2}$ mbar, according to the specific design of the gun. Clearly, a part of the electric energy will get dissipated via extended plasma generation, partial heating and other processes. In what follows we will not discuss these losses and will consider the useful (ablating) part of the energy as proportional to the total energy.

Epitaxial LSMO thin films were deposited on matching substrates: $NdGaO_3$ (110) and $SrTiO_3$ (001) (NGO and STO, respectively); the mismatch is -0.3 % for NGO and 0.9 % for STO. A profilometer was used for thickness calibration.

Films morphology was analysed by means of atomic force microscopy (AFM), the magnetotransport behaviour was characterised by measuring the films resistance and magnetoresistance from 100 K up to 350 K. The magnetotransport measurements were taken as R(H) at different temperatures. The magnetoresistance MR was defined as $(R_{(0)}-R_{(3\ kOe)})/R_{(0)} \cdot 100$. The magnetometry characterization as a function of temperature was carried out by means of Superconducting QUantum Interference Device (SQUID) magnetometry in the temperature range 100-370 K, with an applied magnetic field of 1kOe. Extensive Scanning Tunnelling Microscopy (STM) characterizations fully confirmed the AFM results and obtained clear atomic resolution, these data will be published elsewhere.

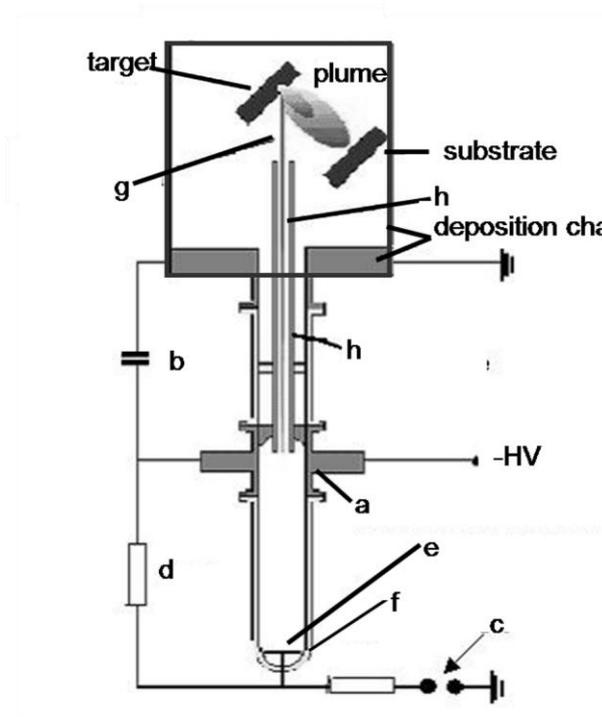

**Figure 1: CSA scheme**
Scheme of the Channel Spark Ablation system. The generator consists of a transient hollow cathode connected to a dielectric acceleration tube and the trigger. The dielectric tube is connected through a narrow exit to the transient hollow cathode and picks up the electron flow for the final acceleration which form the electron beam. The inner wall of the deposition chamber is the actual anode. The trigger circuit consists of the air gap with a floating electrode (c), the charging resistor (d), bottom anode (e).

## 3. Results

LSMO films were grown at substrate temperatures in the range between 700 and 950 °C. The temperature was calibrated by a pyrometer pointing on a black and opaque Nb:STO dummy substrate. The discharge voltage was varied from 7 to 10.5 kV. The pressure was tuned between 2.2 and 4.3 · $10^{-2}$ mbar and actually the accuracy of its detection in the used setup was about 0.1 · $10^{-2}$ mbar (Leybold thermovac). This range enables an efficient CSA processing given its plasma based operation. An additional conditioning is routinely performed on all the samples involving a soft (400°C) annealing in vacuum for a time corresponding to 1.5 · $t_D$, where $t_D$ is deposition time. This procedure was found to increase considerably the reproducibility of the main film parameters and was tentatively ascribed to the removal of the over-oxygenation [20].

### 3.1 Oxygen Pressure

We investigate first the influence of the growth pressure on film morphology. The pressure in the chamber is a very important parameter given the fact that CSA deals with plasma initiation, evolution and propagation. We start from thick films (more than 30 nm), that is films exceeding the thickness where any substrate induced strain is substantially removed. Figure 2 reports the AFM morphology of 45 nm thick films deposited at two different pressures: 2.2 · $10^{-2}$ mbar and 4.2 · $10^{-2}$ mbar. The film deposited at lower pressure has a high roughness $R_{rms}$ above 7 nm and features a highly granular surface with a number of outgrowths. Contrarily, the high pressure regime provides a considerably reduced roughness and $R_{rms}$ values of about 2 nm is achieved. It can be seen thus that the pressure has a crucial influence on the surface roughness both for the presence of outgrowths and particulate.

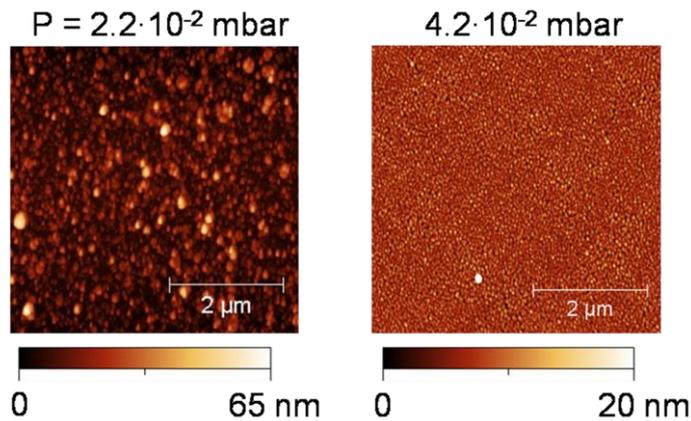

**Figure 2: Chamber pressure effect**
Surface of two 45 nm thick LSMO films deposited at the same condition but the chamber pressure, as indicated. $R_{rms}$ is 7.2 nm for the film deposited at lower pressure and 2.1 nm for the sample deposited at higher pressure.

Considering that the roughness in manganite films is known to be thickness dependent [30], the high enough thickness of the above discussed films (45 nm) does not allow the realization of atomically flat surfaces. This is achieved on the next batch of samples where three 10 nm films have been fabricated while tuning further the working pressure: pressure values of $3.0 \cdot 10^{-2}$ mbar, $3.9 \cdot 10^{-2}$ mbar, and $4.3 \cdot 10^{-2}$ mbar were investigated. The respective morphologies are reported in Fig. 3. The surface clearly becomes cleaner (no outgrowths) and smoother by increasing the pressure but what is more important, the film deposited at $4.3 \cdot 10^{-2}$ mbar shows a roughness confined below 1 nm for a large area of 5 x 5 $\mu m^2$. Such a roughness detected on reasonably large area matches various device constrains allowing the fabrication of atomically flat interfaces in photolithography based microsystems. This surface quality is extended at least up to 20 nm thick films (Figure 7) but the thickness evolution of optimized films is beyond the aim of the presented work.

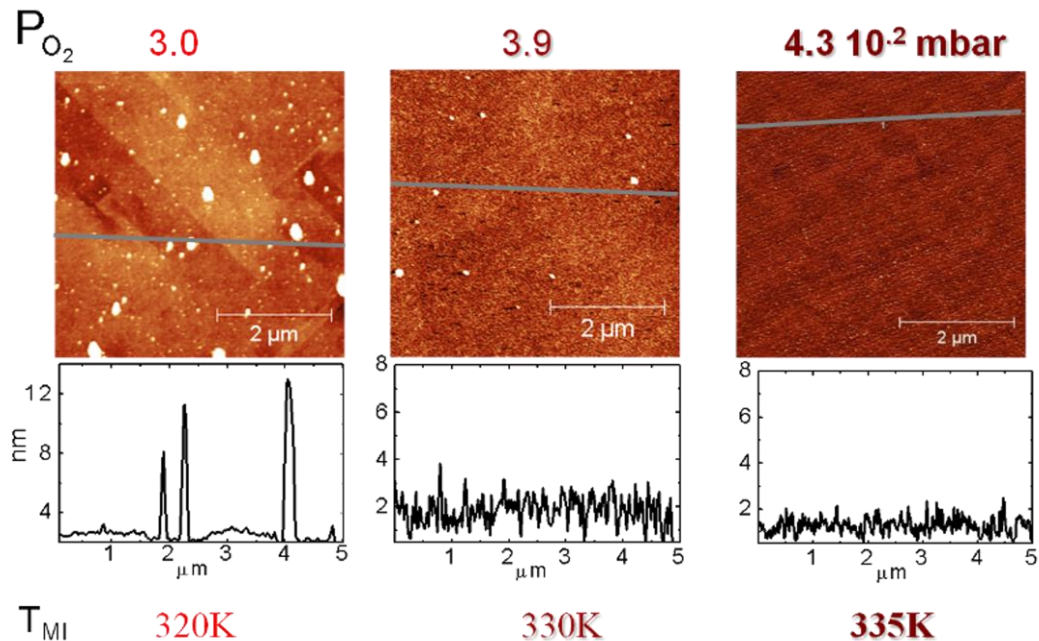

**Figure 3: Pressure driven morphology**
AFM measured morphologies of 10 nm thick LSMO/STO thin films deposited at different pressures. The images are 5 x 5 μm² in size. The relative metal-insulator transition temperature $T_{MI}$ and deposition oxygen pressures ($P_{O_2}$) are reported in the figure.

### 3.2 Pulse energy

The role of the energy accumulated in capacitors (see Fig. 1) and afterwards transferred (partially) into ablation capability has been studied by analyzing separately different accelerating voltages and capacitances. These effects were investigated on 10 nm thick LSMO films deposited at 815 °C at the fixed pressure of $3.9 \cdot 10^{-2}$ mbar.

Figure 4 reports the morphology of LSMO films of 10 ± 1 nm deposited at different voltages ranging from 7 to 10.5 kV on STO and NGO. The main result is the increase of the rms roughness of the film surfaces with voltage for both substrates. Increasing voltage from 7 to 10.5 kV the surface changes from a very smooth state, with an rms roughness below half a unit cell, to a quite harsh one, with rms roughness of almost three unit cells and a peak to valley roughness of several nanometres. Since all the deposition parameters were kept the same, the morphology dependence on the accelerating voltage is ascribed to the ablation.

Optimized parameters for the channel spark deposition of LSMO.

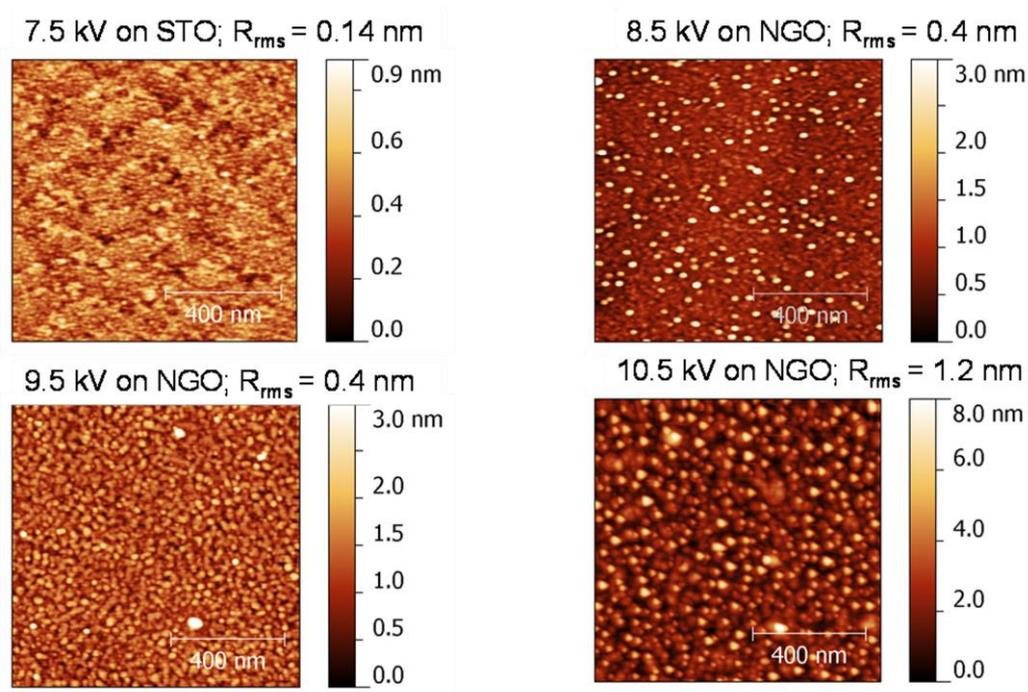

**Figure 4: Voltage effect**
AFM images of 10 nm LSMO films deposited on STO or NGO at different accelerating voltages. The roughness increases with the voltage, i.e. the pulse energy.

The dependence of the particulate density on the capacitance of the CSA has been studied at optimized pressure and voltage. By varying the capacitance from 20 nF (left) to 13.4 nF (right) capacitances no significant variation in the particulate density is detected as function of capacitance (Figure 5). This result is in contrast with other studies claiming the reduction of the particulate density for lower capacitances as a consequence of the decreased beam energy [31-33]. Nevertheless, in those studies the accelerating voltages were much higher and the working pressure was much lower than our optimized conditions.

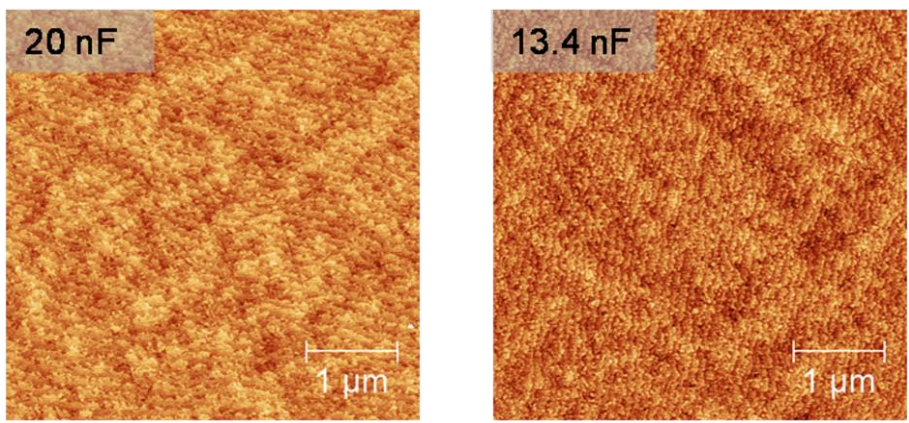

**Figure 5: Capacitance effect**
AFM images of two 10 nm thick LSMO films deposited with a capacitance of 20 nF (right) and 13.4 nF (left) under optimised pressure and voltage conditions. The step-terrace morphology appears identical.

### 3.3 Substrate Temperature

The substrate temperature effect on perovskite thin film morphology has been investigated since the early studies in HTC superconductor films [34] . Very recently studies have been reported also for LSMO where the temperature was identified as a powerful instrument acting on the kinetics of the growth via a considerable diffusion modification [35].

We investigated the temperature dependence of the morphology (Figure 6) on three 10 nm thick films deposited respectively at 950, 850 and 750°C. One can see that at the highest temperature the growth appears dominated by the formation of crystallites with regular shape and facets oriented at 90° each other. This shape remembers the substrate orientations and could correspond to non-stoichiometric aggregates which nucleate on top of the surface [36-38], where the non-stoichiometry could be also represented by local oxygen variations [20]. Decreasing temperature the surface becomes smoother and at 750 °C one can see terraces coming from the substrate as a signature of the possible step flow growth.

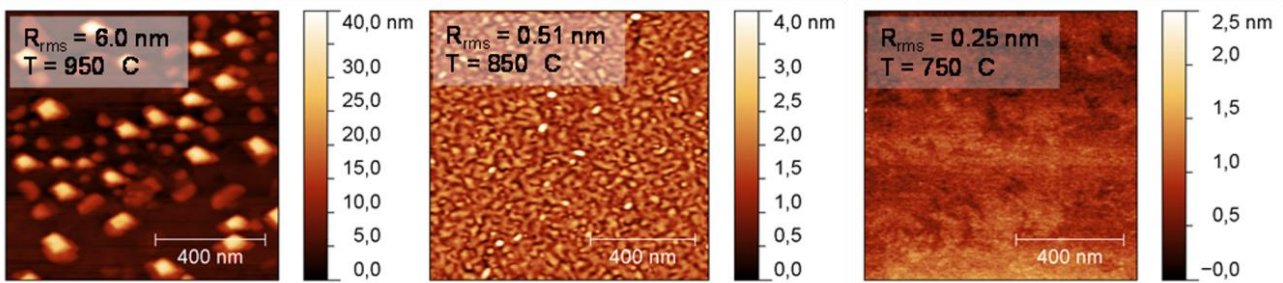

**Figure 6: Temperature effect**
AFM images of samples grown at different substrate temperatures, as indicated. Decreasing substrate temperature the surface becomes smoother and smoother.

### 3.4 Magnetic characterizations

The optimization of the surface morphology will lead to a real enhancement of LSMO performances only if magnetic properties of such optimized films would remain at least as robust as for rougher films. We start magnetic investigations from the SQUID magnetometry of a most standard optimised 20 nm thick film deposited on STO (Figure 7). The sample was zero field cooled and the magnetic moment was detected by keeping an applied field of 1 kOe. The film presents high magnetic moment and a $T_C$ of about 325 K. The 325 K reproduces exactly previous (rough morphology) $T_C$ for such thicknesses [9]. The magnetic moment higher than 400 emu/cm$^3$ at 100 K is consistent with bulk sample values [39].

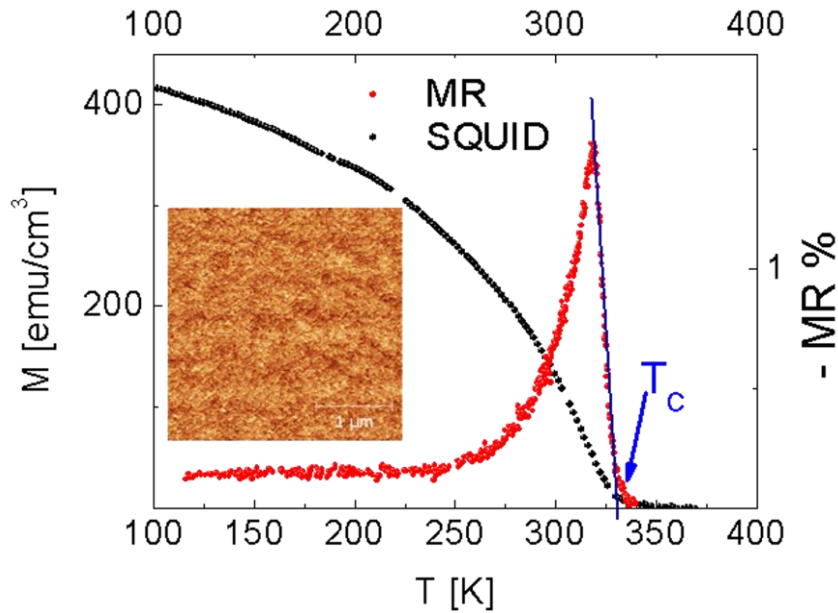

**Figure 7: SQUID and magnetoresistance comparison**
Comparison between SQUID (black) and magnetoresistance (red) for a typical 20 nm LSMO film. The $T_C$ can be defined also from magnetotransport measurement. The inset shows an AFM image, size 9 μm$^2$, the maximum peak-valley roughness is 1 nm and the rms roughness is 0.26 nm.

Figure 7 reports also the magnetoresistance measurements of *the same* film. As typical for manganites the peak of the MR is placed at temperatures slightly below the Curie point confirming that the identification of the magnetoresistance peak with the Curie point is not correct [40]. More interestingly, the critical temperature coincides exactly with the linear extrapolation to zero of the MR part right to the peak. It looks appealing thus to advance the magnetoresistance MR(T) measurement as an appropriate and an extremely simple method for the $T_C$ definition via magnetotransport. This appears to be especially useful for ultrathin films, where most standard magnetic techniques could fail. The inset in Figure 7 reports a typical AFM image for a 20 nm thick film showing a rms roughness of 0.26 nm and a maximum height of the roughness of 1 nm.

The shape of magnetoresistance with magnetic field and at different temperatures is shown in Figure 8. It shows normalized R(H) curves for a 10 nm film deposited at 7 KV on NGO, measured at different temperatures in the range ± 288 mT. The magnetoresistance (MR) increases by increasing temperature in the ferromagnetic region. At a certain temperature the MR has a peak, then decreases (see the inset). This peak corresponds with a change in the shape of the R(H) curve [39, 41], from cuspide-like, black downward triangles in figure 8, to calotta-like, red upward triangles in figure 8. The inset shows the temperature dependence of the magnetoresistance value.

The MR in the range ± 100 mT corresponds to the so called Low Field Magneto-Resistance (LFMR): such low MR values as those presented in figure 8 generally indicate a high epitaxial quality of the film, while the presence of grain boundaries would increase LFMR up to several per cents [40]. Indeed, ideal epitaxial LSMO films have very low LFMR, less than 1 % in the low T region [42]. In Figure 8 is possible to observe that the LFMR is less than 0.5% at 100 K and remains below 2 % in the whole temperature range.

Figure 9 communicates an extremely important result of this paper – the transport and magnetotransport properties of these optimized films are weakly dependent on the morphology. We show this on the basis of a set of films grown at different voltages (and hence different morphologies, see Figure 4), but the property holds also for other growth parameters variations. The temperature dependence of resistance and MR for several 10 nm films is reported in Figure 9, where the sample from Figure 8 corresponds to blue data set. Both MR(T) (full circles) and the R(T) (open circles) of four 10 nm thick films deposited at indicated voltages are compared. The R(T) curves display the resistance at zero magnetic field while the MR(T) curves report the resistance change between 0 Oe and ± 288 mT, where each data point is obtained by measuring R(H) at a constant temperature. One can see that R(T) curves are nearly identical while the variation of the MR(T) peaks could be ascribed to the possible inaccuracy in the thickness definition (rather than to morphology variation). The lower values of MR for the films deposited on NGO can be ascribed to the better substrate-film lattice matching. It would be interesting to compare the magnetoresistive properties with literature data for manganite films of similar thickness. Unfortunately similar characterizations for this range of fields and thicknesses are not available in the literature. The only data for very thin films features several percents of MR but are measured at 2.5 T and R(H) curves are not shown [24]. The comparison with thicker films reveals that the magnetoresistive behaviour of our films is comparable with the epitaxial ones [40, 42].

The list of the optimized parameters ranges are reported in table I. The effect of the substrate temperature on the magnetotransport properties will require further work.

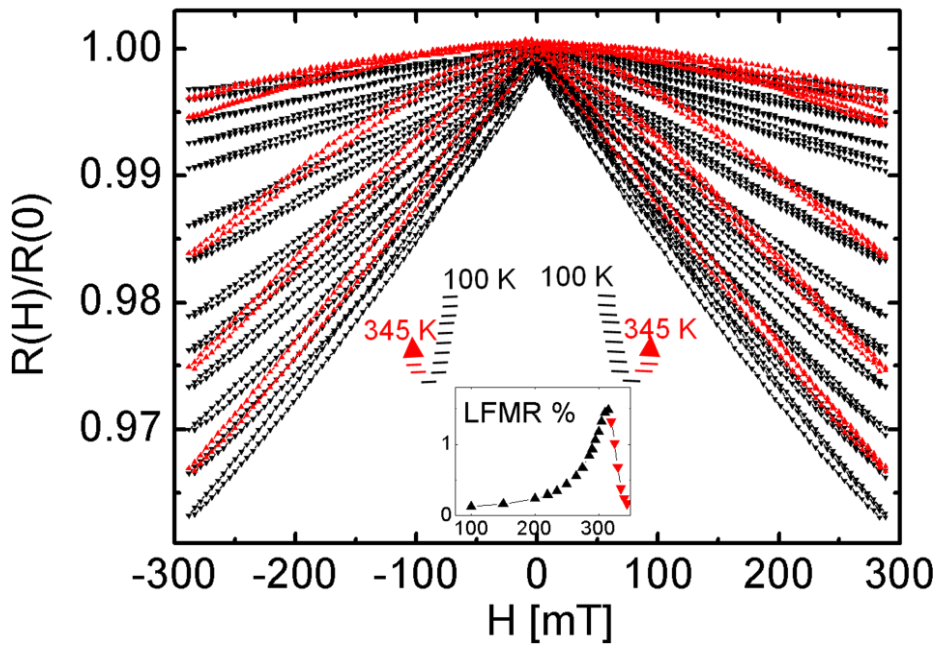

**Figure 8: Magnetoresistance of a selected 10 nm film**
R(H) at constant temperatures, normalized at the value at zero field, for a 10 nm thick LSMO film on NGO (110). The black curves are descending while the red ones are asceding. Also the shape of the curves changes when the temperature trend is inverted. The inset reports the temperature trend of the LFMR ( ± 100mT ) amplitude.

**Table I : Optimized parameters**
Optimized parameters for the channel spark deposition of LSMO.

| $P_{O2}$ [$10^{-2}$ mbar] | V [kV] | C [nF] | $T_S$ [°C] |
|---|---|---|---|
| | | | |
| 3.5 to 4.5 | 7 to 8 | Not affecting | 750 to 800 |

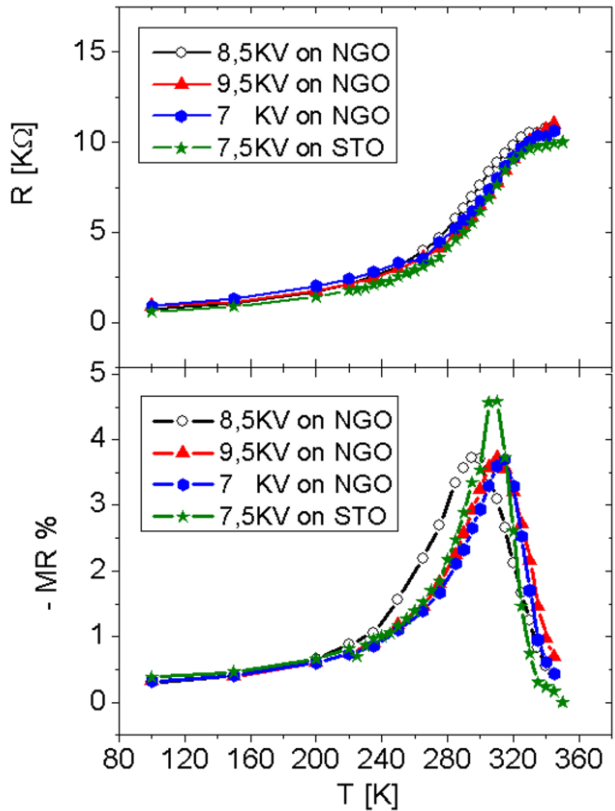

**Figure 9: Magnetotransport of 10 nm LSMO films**

MR(T) and R(T) of 10 ± 1 nm thick films deposited at different voltages, as indicated.

## 4. Discussion

The straightforward role of the pressure in diminishing the particulate density on the film surface is enlightened by the ablation concepts developed for PLD. First we have to note that the problem of particulate coming from the target is inherent in the ablation processes [43, 44]. Commonly one observes two classes of particulate roughly distinguished in over-micron round shaped particles and submicron round or faceted particulate [45, 46]. The bigger particles are thought to come from the target as a result of microcracks, pits and weakly attached particles. The submicron particles are due to the superheating of the subsurface region [46], which reaches a higher temperature than the surface, cooled by the latent heat of evaporation. In this latter case the gaseous phases nucleate and cause volume expansion which leads to explosive sublimation. In the superheated regions target material segregates and after some pulses is ejected in form of submicron particles [47], known to have a different chemical composition from the target as observed in HTC [36, 47] and LSMO [37, 38].

In order to avoid sub-micron particulate due to subsurface heating effect one has to use very dense (94-97% of the theoretical value) and freshly polished targets [36, 45]. Since the subsurface superheating model describes that the superheated region depth increases with pulse energy and duration, one should work just above the ablation threshold and use short pulses [46, 48]. Also the penetration depth does matter: less it is, less deep the superheated region, with a strong improvement in the morphology [43, 48].

In channel spark discharge it is known that increasing the background gas pressure the electrons beam energy distribution is shifted towards lower values, the beam power decreases and the pulse duration is shorter [49, 50]. Thus by tuning one parameter we are moving towards the minimization of the subsurface superheating. The above argument enables us to rationalize not only our data but also qualitatively similar, but not commented, results found in pulsed electron ablation of HTC precursor [31].

The role voltage dependence of the roughness is explained by the fact that increasing the accelerating voltage the ablation becomes strenger and hence the rate per pulse is higher. This observation fits well with the Venables model for the sub-monolayer epitaxial growth [51]. Provided the density of stable nuclei, having size bigger than the critical one $i$ (the minimum size in order to avoid the dissociation of the nucleus), is $dn_x/dt = \sigma_i Dn_i R/\tau_a + \sigma_x Dn_x R/\tau_a + RZ$, where $\sigma_i$ and $\sigma_x$ are rispectively the capture numbers for critical and stable nuclei, $n_i$ and $n_x$ are the surface density of critical and stable nuclei, $D$ the adatoms diffusivity, $R$ the arrival rate of adatoms and $\tau_a$ their desorption time, $Z$ the covered surface ratio, it appears that the increased flux of adatoms, consequence of a stronger ablation, is thought to be responsible for the rapid aggregation and the increased roughness.

It is our opinion that the capacitance does not have a straightforward role for the investigated ranges of pressure and voltage. This can be rationalised via simple qualitative considerations. The energy stored in the capacitors is $\frac{1}{2} \cdot C \cdot U^2$, where $C$ is the capacitance and $U$ the accelerating voltage. The quadratic dependence makes $U$ more influent than $C$. Moreover decreasing the capacitance also the $RC$ constant is decreased, the stored energy is transferred in a shorter time so the power transferred to the target remains the same. This argument can explain why, once pressure and voltage are optimized, the capacitance plays a minor role.

The fact that bidimensional growth is stabilized at relatively low temperature was found in LSMO also by Bachelet *et al.* [35]. The interpretation relies on the fact that at higher temperature (higher diffusivity) adatoms can move longer on the surface and eventually find the minimum energy position. This does not imply necessarily atomically flatness but rather regular atoms positioning, which can as well arrange in three dimensional way. Decreasing substrate temperature the surface diffusion is decreased and the bidimensional step flow is preferred to 3D nuclea, provided the substrate temperature is still sufficient to ensure a diffusion length comparable with the terrace length $\lambda_D=\sqrt{(D_s\tau)} \sim 400$ nm for STO substrate. The terrace sides are kinks, i.e. points with higher density of dangly bonds and consequently represent sites of preferential nucleation. If the temperature is too low ($\lambda_D<<\sqrt{(D_s\tau)}$) the adatoms cannot reach the kinks, resulting in a rough 3D growth with poor epitaxy. Oppositely, for too high substrate temperatures, the adatoms have sufficient energy to detach from the kink and to form stable nuclei also in the middle of the terrace (also 3D like growth tendency).

All the observations are summarized in the fact that the whole film growth is completely determined by the energy density transferred from the beam to the target and by the substrate temperature. The control of these two factors enables the control of the morphology and magnetic and transport properties of the CSA deposited manganite thin films.

## 5. Conclusions

In this work we have defined and described a robust set of protocols for the growth of low roughness (rms roughness below the unit cell size) LSMO manganite thin films by means of the Channel Spark Ablation technique, a cheap and home building setting. This surface smoothening did not modify the magnetic properties, which remained robust and in line with literature data for similar thicknesses. On the basis of the performed magnetotransport analysis we propose as a very versatile and simple method the definition of the $T_C$ from the magnetoresistance slope. Presented results confirm that CSA is fully competitive with the more wide spread PLD technique for the growth of high quality manganite thin films.

### Aknowledgements

The authors thank Federico Bona for technical help. The extensive use of the scanning probe microscope at the "Centro Interfacoltà Misure" of the University of Parma is acknowledged. The authors also acknowledge the financial support from the FP7 project NMP3-LA-2010-246102 (IFOX) and from NMP-2010-SMALL-4-263104 (HINTS).